\magnification=1200
\hsize=15.6 truecm
\vsize=23.6 truecm

\centerline{\bf DYNAMIC INSTABILITY CRITERION FOR }
\centerline{\bf CIRCULAR (VORTON) STRING LOOPS}
\vskip 1 cm
\centerline{\bf Brandon Carter and Xavier Martin}
\vskip 0.4 cm
\centerline{\it D\' epartement d'Astrophysique Relativiste et de
Cosmologie, C.N.R.S,}
\centerline{\it Observatoire de Paris, 92 195 Meudon, France.}
\vskip 0.4 cm
\centerline{March, 1993.}
\vskip 1 cm

\noindent
{\bf Abstract.} Dynamic perturbation equations are derived for a generic 
stationary state of an elastic string model -- of the kind appropriate for 
representing a superconducting cosmic string -- in a flat background. In the 
case of a circular equilibrium (i.e. vorton) state of a closed string
loop it is shown that the fundamental axisymmetric ($n=0$) and lowest
order ($n=1$) nonaxisymmetric perturbation modes can never be unstable.
However, stability for modes of higher order ($n\geq 2$) is found to be
non-trivially dependent on the values of the characteristic propagation 
velocity, $c$ say, of longitudinal perturbations and of the corresponding 
extrinsic perturbation velocity, $v$ say. For each mode number the criterion 
for instability is the existence of nonreal roots for a certain cubic 
eigenvalue equation for the corresponding mode frequency. A very simple
sufficient but not necessary condition for reality of the roots and therefore
absence of instability is that the characteristic velocity ratio, $c/v$
be greater than or equal to unity. Closer examination of the low velocity
(experimentally accessible) nonrelativistic regime shows that in that limit
the criterion for instability is just that the dimensionless characteristic 
ratio $c/v$ be less than a critical value $\chi_c$ whose numerical value is
approximately $1\over 2$. In the relativistic regime that is relevant to 
superconducting cosmic strings the situation is rather delicate, calling 
for more detailed investigation that is postponed for future work.
\vskip 1 cm

\bigskip\noindent
{\bf 1. The fundamental dynamical equations of an elastic
string model}
\medskip

The ultimate motivation of the present work is to deal with a problem
that was implicitly raised when it was first pointed out by Davis and
Shellard [1-3] that superconducting string loops might be ``stabilised''
(as ``vortons'') by centrifugal force (which can be expected to be very 
much stronger than the inefficient magnetic ``spring'' repulsion effect 
that had been previously considered by some authors [4-7]) with the 
implication that (unlike nonconducting cosmic string loops) such current 
carrying loops might survive long enough to produce a cosmological mass 
excess. However, although Davis and Shellard provided the first examples 
of a general category of circular rotating equilibrium states, their use of 
the term ``stabilisation'' was rather premature at that stage because they 
had not carried out any verification of the actual stability of such 
equilibria with respect to ordinary macroscopic dynamical perturbations.
(What they did do was to consider the range of parameters for which the
equilibrium would be able to survive against macroscopic quantum tunnelling.) 
It was in fact fairly easy to confirm that the equilibrium states in 
question would be stable against strictly axisymmetric (circularity 
conserving) perturbations [8, 9] but the present works constitutes what,
as far as we know, is the first investigation of stability of centrifugally
supported string loops against general nonaxisymmetric perturbations. The
range of the present study is not limited to the strongly relativistic
regime relevant to applications to cosmic strings (which will be 
considerered in more detail in a following article) but covers quite 
general elastic string models including those relevant to the 
experimentally accessible nonrelativistic limit.

When we first set out on this investigation we had expected that its
outcome would be expressible simply as a general stability theorem --
confirming for generic dynamical perturbations what we already knew to be 
the case for axisymmetric ones. However, the picture that actually emerges
is less simple and more interesting. It turns out that, whereas for many 
kinds of string model the circular equilibrium states will indeed always be
stable, nevertheless there are other models for which some or even all of 
the circular equilibrium states are actually unstable. The domain of
instability is not restricted to the strongly relativistic regime 
relevant for superconducting cosmic string theory and its cosmological
implications (which, as remarked above, will be considered in more detail
in a subsequent article) but it extends down into the Newtonian regime.
What will be confirmed, however, is that the requirements for dynamic 
stability will always be satisfied in the Hookean (low tension) limit.

The plan of this work is as follows. After completing this introductory
section by summarising the general principles governing the dynamics of the
category of elastic string models under consideration, we go on to provide,
in Section 2, a brief derivation of the conditions for stationary 
equilibrium with repect to a nonrotating frame in a flat background. Section 
3 contains a derivation of the local dynamic equations governing the
evolution of small perturbations of such equilibrium states, ans Section 4 
describes the way to obtain global solutions of these equations for cases
in which the unperturbed equilibrium states are circular (the study of more
general equilibrium states being left for subsequent work).

Any two-dimensional worldsheet can be conveniently described [8,10] in
terms of some chosen orthonormal frame consisting of mutually orthogonal
timelike and spacelike tangent vectors $u^\mu$ and $v^\mu$ as  
characterised by
$$u^\mu u_\mu=-1\ ,\hskip 1 cm u^\mu v_\mu=0\, ,\hskip 1 cm
v^\mu v_\mu=1\, ,\eqno{(1.1)}$$
on the understanding that we are using units such that the speed of light
is set to unity. In particular, any such frame can be used to provide an
explicit representation of the antisymmetric tangent element tensor of
the surface, which will be expressible as
$${\cal E}^{\mu\nu}=u^\mu v^\nu- u^\nu v^\mu\, ,\eqno{(1.2)}$$
and which is well defined -- independently of the choice of frame --
modulo an ambiguity of sign which is resolved by a choice of orientation in 
the surface. The square of this surface element tensor gives the first
fundamental tensor of the worldsheet as
$$ \eta^{\mu\nu}={\cal E}^{\mu\rho}{\cal E}_\rho^{\ \nu}=
-u^\mu u^\nu+ v^\mu v^\nu\, ,\eqno{(1.3)}$$
which is geometrically well defined, without any ambiguity at all, as
also is the second fundamental tensor $K_{\mu\nu}^{\ \ \rho}$, which is
defined (for an embedded surface of arbitrary dimension) in terms of the
corresponding first fundamental tensor by the formula 
$$K{_{\mu\nu}}^{\!\rho}=\eta{_\mu}^{\!\sigma}\tilde\nabla_{\!\nu}
\eta{_\sigma}^{\!\rho}\, ,\eqno{(1.4)}$$
where the surface tangential covariant differentiation operator is
defined by
$$\tilde\nabla_{\!\mu}= \eta{_\mu}^{\!\nu}\nabla_{\!\nu}\, .\eqno{(1.5)}$$
As well as the obvious orthogonality and tangentiality properties
$$K{_{\mu\nu}}^{\!\sigma}\eta{_\sigma}^{\!\rho}= 0\, ,\hskip 1 cm
\perp^{\!\mu}_{\,\sigma}K{_{\mu\nu}}^{\!\rho}=0\, ,\eqno{(1.6)}$$
where the orthogonal projection operator is defined by
$$\perp^{\!\mu}_{\,\nu}=g^\mu_{\,\nu}-\eta^\mu_{\,\nu}\, ,\eqno{(1.7)}$$
the second fundamental tensor has the nontrivial Weingarten symmetry
property [8, 11, 12] expressible -- using square brackets to denote index 
antisymmetrisation -- as
$$K{_{[\mu\nu]}}^{\rho}=0\, .\eqno{(1.8)}$$

As in the preceding investigation of local stability conditions and
characteristic speeds [8, 11], condition (1.8) will play an important role
in the calculation that follows, since although it will hold automatically
as an identity if $K{_{\mu\nu}}^{\!\rho}$ is derived from a two-surface
that is already specified, on the other hand it must be deliberately 
imposed as an indispensible (Frobenius type) integrability condition if
one chooses, as we shall do in this work, to use an approach in which the
surface is considered as a secondary construct determined implicitly by 
prior specification of the surface element ${\cal E}^{\mu\nu}$.
Our reason for using such an approach -- i.e. for considering the world
sheet to be constructed from the tangent tensor  ${\cal E}^{\mu\nu}$
rather than the other way round -- is that, although our ultimate purpos is 
to consider the world sheet of a single isolated string loop, it will be 
computationally convenient to consider it as a member of a fictitious 
space-filling congruence of string loops, each separately subject to the 
same dynamical laws. This will enable us to treat quantities such as 
${\cal E}^{\mu\nu}$ as ordinary space-filling tensor fields free from the
restriction of having their support restricted to any single world sheet.
This facilitates the use of an economically covariant formalism that
avoids the need to become involved in the questions of choice of gauge that 
would arise if we used an approach of the more traditional kind involving
the explicit specification of the string worldsheet in some particular
coordinate gauge.

In addition to the purely geometrical evolution condition (1.8) the
calculation that follows will depend on the use of appropriate extrinsic and
internal dynamical equations. For any classical string model -- in the limit
in which finite thickness effects including polarisation are neglected --
the extrinsic dynamical equation of motion governing the motion of the world
sheet will be given, whenever external forces are absent as is postulated in 
the present work, by a ``generalised sail equation'' [8, 11, 12] of the form
$$\tilde T{^{\mu\nu}} K{_{\mu\nu}}^{\!\rho}=0\, ,\eqno{(1.9)}$$
where $\tilde T{^{\mu\nu}}$ is the relevant (surface) stress momentum energy 
density tensor, whose specification will depend on the kind of string model 
under consideration. (For example, in the well known degenerate cas of a
Goto-Nambu string model the tensor  $\tilde T{^{\mu\nu}}$ is simply
proportional to the fundamental tensor $\eta^{\mu\nu}$.

There is no loss of generality in considering $\tilde T{^{\mu\nu}}$ to be
specified by its eigenvalues, $T$ and $U$ say, which are interpretable 
respectively as the intrinsic string tension and as the energy density in
the preferred rest frame determined by its spacelike and timelike 
eigenvalues $v^\mu$ and $u^\mu$ which can conveniently be used to specify
the choice of the frame introduced in (1.1). The surface stress momentum
energy density is thus expressible in the generic form
$$\tilde T{^{\mu\nu}}= U u^\mu u^\nu-T v^\mu v^\nu\, ,\eqno{(1.10)}$$
where what distinguishes one kind of string model from another are the
rules specifying the energy density $U$ and the tension $T$.

Although more complicated possibilities can be envisaged (and would be
necessary for dealing with cases in which internal dissipation is
important), our present study will be limited to the category of simple 
elastic string models, in which the rules for specifying $U$ and $T$ consist 
just of a single equation of state specifying one of them as a monotonic 
function of the other. (The Goto-Nambu case can be classified within this
category as the degenerate limit case in which $U$ and $T$ are fixed 
with the same value $U=T=m^2$, where $m$ is a constant with the 
dimensionality of mass in units of the kind used in the present work for
which the speed of light is set to unity.) As well as providing a good
description for such familiar examples as ordinary violin strings, such 
simple elastic string models are also [8, 9, 13] appropriate for the more
exotic case of superconducting cosmic string models whose investigation
provided the primary motivation for this work and which are adequately
described by the formalism used here, provided that electromagnetic
coupling corrections (of the kind investigated by Peter [14, 15] are
unimportant -- as will commonly be the case in the cosmological
applications we have in mind [1-3, 16, 17].

Whatever the equation of state may be, the internal equations of motion
for such a simple elastic string model can be shown [8, 9] to reduce just
to a pair of surface current conservation laws expressible as
$$ \tilde\nabla_{\!\rho}\big(\nu u^\rho\big)=0\, ,\eqno{(1.11)}$$
and
$$ \tilde\nabla_{\!\rho}\big(\mu v^\rho\big)=0\, ,\eqno{(1.12)}$$
where $\nu$ and $\mu$ are a number density variable and an associated
effective mass variable -- or chemical potential -- which are determined by
the equation of state. It is convenient to consider the equation of state 
to be specified by the expression for the energy density $U$ as a function 
of the number density $\nu$.  The corresponding function for the chemical
potential $\mu$ will then be given as the derivative
$$ \mu= {\rm d}U/{\rm d}\nu\, , \eqno{(1.13)}$$
while finally the function for the string tension $T$ will be given by
$$ T=U-\nu\mu\, .\eqno{(1.14]}$$

It is to be remarked for reference later on that the number consrvation law
(1.11) can be expressed in the equivalent adjoint form
$${\cal E}^{\rho\sigma}\nabla_{\!\rho}\big(\nu v_\sigma\big)=0\, ,
\eqno{(1.15})$$
while the analogous irrotationality condition obtained as the adjoint of
(1.12) is
$${\cal E}^{\rho\sigma}\nabla_{\!\rho}\big(\mu u_\sigma\big)=0\, .
\eqno{(1.15})$$
In terms of the preferred reference frame introduced by (1.10) the extrinsic
equation of motion (1.9) for the world sheet can be written in the more
explicit form
$$\perp^{\!\mu}_{\,\rho}\big( U u^\nu\nabla_{\!\nu} u^\rho - T
v^\nu\nabla_{\!\nu} v^\rho\big)=0\, , \eqno{(1.17)}$$
while, finally, the geometric integrability condition (1.8) will
take the form 
$$\perp^{\!\mu}_{\,\rho}\big( u^\nu\nabla_{\!\nu} v^\rho -
v^\nu\nabla_{\!\nu} u^\rho\big)=0\, . \eqno{(1.18)}$$

\bigskip\noindent
{\bf 2. Equilibrium with respect to a static Killing vector in a flat
background}
\medskip

The concept of equilibrium is generally understood to mean invariance with
respect to an at least locally timelike Killing vector $k^\mu$ of the 
relevant background spacetime, where the Killing condition for the 
invariance of the background spacetime metric $g_{\mu\nu}$ itself is
expressible in terms of Riemannian covariant differentiation as
$$\nabla_{\!(\mu} k_{\nu)}=0\, ,\eqno{(2.1)}$$
using round brackets to indicate symmetrisation. The Killing vector is
said to be {\it static} in the special case in which it also satisfies the
irrotationality condition 
$$k_{[\mu}\nabla_{\!\nu} k_{\rho]}=0\, .\eqno{(2.2)}$$
In the case of a flat background spacetime, the combination of the Killing
condition (2.1) and the staticity condition (2.2) is expressible simply as
the condition of being covariantly constant, i.e.
$$\nabla_{\!\mu} k_\nu=0\, ,\eqno{(2.3)}$$
which is the equation characterising a simple linear translation. Subject to
the appropriate signature convention, we assure that the generator $k^\mu$ 
is timelike by imposing the usual normalisation condition
$$ k^\mu k_\mu=-1\, .\eqno{(2.4)}$$

The requirement that the string should be in equilkibrium with respect to
$k^\mu$ includes as a first -- purely topological -- condition the
requirement that $k^\mu$ should be tangential to its world sheet. It follows
that at each point in the world sheet it will be possible to choose an 
orthogonal spacelike unit tangent vector $e^\mu$ say, as characterised by
$$ e^\mu e_\mu=1\, ,\hskip 1 cm e^\mu k_\mu=0\, ,\eqno{(2.5)}$$
in terms of which the fundamental tensor of the world sheet will be
expressible as
$$\eta^{\mu\nu}= -k^\mu k^\nu +e^\mu e^\nu\, ,\eqno{(2.6)}$$
where $e^\mu$ is uniquely determined by a single global choice of 
orientation on the (topologically cylindrical) world sheet of the 
string loop.

The condiion of equilibrium with respect to a vector field $k^\mu$ means that 
any physically well defined tensor field on the world sheet should be 
invariant under its action. With respect to a vector $k^\mu$ satisfying 
the covariant constancy condition (2.3) (which implies that the corresponding
Lie differentation operation reduces to one of direct covariant 
differentiation) the condition of invariance simply means that the
corresponding covariant derivative should vanish. Thus in the case of the
spacelike unit vector $e^\mu$ the invariance condition is simply
$$k^\nu\nabla_{\!\nu} e^\mu=0\, .\eqno{(2.7)}$$
Using this in conjunction with (2.3) we can immediately evaluate the
second fundamental tensor (1.4) of the world sheet as
$$K{_{\mu\nu}}^{\!\rho}=e_\mu e_\nu K^\rho\, ,\eqno{(2.8)}$$
where its trace, the curvature vector $K^\rho$, is given by
$$ K^\rho =K^{\nu\ \rho}_{\ \nu}= e^\nu\nabla_{\!\nu} e^\rho\, .
\eqno{(2.9)}$$

Although we are restricting ourselves to cases of equilibrium with respect
to a Killing vector that is static (i.e. the present work does not include
consideration of equilibrium with respect to a rotating frame) we cannot
impose the requirement that the equilibrium itself be static in the 
strictest sense -- which would mean that the intrinsically preferred unit
vector $u^\mu$ of the string would be aligned with the Killing vector 
generator $k^\mu$ -- since (as shown by previous more general studies 
[18, 19]) such a strictly static  equilibrium is possible only for states in 
which the string is actually {\it straight}, meaning that $K^\mu$ should 
vanish everywhere along it, a condition which is obviously incompatible 
with the closed loop topology under consideration here. The kind of
equilibrium states to be investigated here are not strictly static but 
merely stationary, which means they will be characterised by a {\it nonzero}
``running velocity'', $v$ say, expressing the deviation of the 
intrinsically preferred timelike unit stress momentum energy eigenvector 
$u^\mu$ of the string from the Killing vector $k^\mu$, so that it will be
expressible in the form
$$ u^\mu=\gamma\big(k^\mu+ v e^\mu\big)\, ,\eqno{(2.10)}$$
while similarly (subject to the appropriate sign convention) the
corresponding spacelike unit stress momentum energy eigenvector $v^\mu$
will be expressible in the form 
$$ v^\mu=\gamma\big(v k^\mu+ e^\mu\big)\, ,\eqno{(2.11)}$$
where $\gamma$ is the relevant Lorentz factor, i.e.
$$ \gamma=1/\sqrt{1-v^2}\, .\eqno{(2.12)}$$
It follows from the results of the previous paragraph that the derivatives 
appearing in the integrability condition (1.18) will be expressible as
$$ u^\nu\nabla_{\!\nu}v^\mu=v^\nu\nabla_{\!\nu} u^\mu=
v \gamma^2 K^\mu\, ,\eqno{(2.13)}$$
while those appearing in the equation of extrinsic motion (1.17) will be
given by
$$ u^\nu\nabla_{\!\nu}u^\mu=v^2\gamma^2 K^\mu\, ,\hskip 1 cm 
v^\nu\nabla_{\!\nu}v^\mu=\gamma^2 K^\mu\, .\eqno{(2.14)}$$
Substituting this in (1.17) -- or by direct substitution of (2.8) in (1.9)
-- it can be seen that the extrinsic dynamical equation will reduce simply
to
$$ \big(U v^2-T\big) K^\mu=0\, .\eqno{(2.15)}$$
In so far as the longitudinal dynamical equations are concerned, it can be
seen that the combination of the irrotationality condition (1.16) with the
equilibrium requirement that derivatives with repect to $k^\mu$ should
vanish leads to a conservation law of the form
$$ \tilde\nabla_{\!\rho}\big(\gamma\mu\big)=0\, ,\eqno{(2.16)}$$
which is the string analogue of the well known Bernoulli conservation law
in irrotational perfect fluid theory. In the string case (unlike the perfect
fluid case) one can also derive a corresponding dual conservation law from
the other longitudinal dynamical equation (1.15), which similarly leads to
$$ \tilde\nabla_{\!\rho}\big(v\gamma\nu\big)=0\, .\eqno{(2.17)}$$
For a generic equation of state (the noteworthy exception being the
special case of the integrable equation of state characterised by the
condition that the product of $U$ and $T$ be constant [8, 19, 21]) the
constancy conditions (2.15) and (2.17) are independent and therefore
sufficient to determine the constancy of the intrinsic state of the string, 
as specified by $\nu$, i.e.
$$\tilde\nabla_{\!\rho}\nu=0\, .\eqno{(2.18)}$$
and also of the running velocity $v$.

It can be seen from the extrinsic equilibrium condition (2.15) that there
are two qualitatively different ways in which local equilibrium with respect 
to a static Killing vector in a flat background can be achieved. One 
possibiity is for the string to be straight, $K^\mu=0$, in which case the 
running velocity $v$ can have any (subliminal) value, including the zero 
value that correponds to the strictly static case. The other possibility --
the only one that is relevant when the curvature $K^\rho$ is 
nonzero as must be the case for the closed loops with which the present work 
is concerned -- is that of stationary but nonstatic states in which the
running velocity has the nonzero value determined uniquely (modulo a sign)
as a function of the intrinsic string state by the formula
$$ v^2=T/U\, ,\eqno{(2.19)}$$
(which must hold not only in the generic case for which, according to
(2.18) both the running velocity $v$ and the intrinsic string state, as 
characterised by $T$ and $U$ are constant, but also in the exceptional case 
of the integrable string model[19] for which the running velocity $v$ and 
the intrinsic state may vary along the length of the string).

The formula (2.19) is the same as the formula [8, 11] for the characteristic
speed (relative to the intrinsically preferred rest frame specified by 
$u^\mu$) of extrinsic perturbations of the world sheet. The result that 
has just been derived is therefore expressible as the {\it theorem} [18, 19] 
that in any curved equilibrium state of a simple elastic string model with 
respect to a static (non-rotating) Killing vector in a flat background the 
running velocity must be just such as to cancel the characteristic speed of 
relatively {\it backward moving extrinsic perturbations}, which are therefore
static in the sense that the relative velocity, $v_{-}$ say, of such
relatively backward moving perturbations with respect to the static Killing
vector $k^\mu$ is zero, i.e.
$$v_{-}=0\, .\eqno{2.20}$$
It follows that, with respect to $k^\mu$, the speed, $v_{+}$ say, of 
relatively {\it forward moving extrinsic perturbations} will be 
approximately twice the relative runing speed $v$ in the Newtonian limit 
in which the relative speeds are small compared with the speed of light 
(which is unity in the system we are using), the exact expression that will 
be valid for arbitrarily high velocities being given by
$$ v_{+}={2v\over 1+v^2}\, .\eqno{(2.21)}$$

We remark that the relative characteristic speed, $c$ say, of longitudinal
perturbations, which is given [8, 11] by
$$c^2=-{{\rm d}T\over{\rm d}U}={\nu\over\mu}{{\rm d}\mu\over{\rm d}\nu}
\, ,\eqno{(2.22)}$$
may, depending on the equation of state, be less than or greater than or
equal to that of the extrinsic (transverse) perturbations as given by the 
formula (2.19) for $v$ and that in the former case the relatively backward
moving longitudinal perturbations would be absolutely forward moving 
relative to the background frame specified by $k^\mu$. Both from the point
of view of secular stability with respect to radiation backreaction (whose 
analysis, to which we have alluded elsewhere [12, 19] is beyond the scope of 
the present discussion) and, as will be seen below, from the point of view
of ordinary dynamical stability, it is the latter alternative $c\geq v$, 
that is most favorable. As well as applying to the familiar nonrelativistic
Hookean (low tension) limit, the strict inequality $c>v$ will hold in
particular for an equation of state of the simple constant trace type (i.e. 
a constant value for the sum $U+T$) that was implicitly used in many of the
early discussions of superconducting cosmic strings [22-25], for which the 
longitudinal characteristic speed is actually equal to that of light, i.e. 
$c=1$. However, closer examination of the underlying superconducting string
teory by Peter [13] has shown that it would be more accurate to use an 
equation of state for which the longitudinal characteristic speed is in the 
dangerous regime, meaning that it is lower than the transverse 
characteristic speed, i.e. $c< v$, while the special integrable string model
referred to above [20, 21] corresponds to te intermediate case $c=v$.

\bigskip\noindent
{\bf 3. Local perturbation equations for acurved equilibrium state.}
\medskip

 Before proceeding,we need to introduce a mathematical device that is 
frequently useful [11] for dealing with dimensionally restricted physical 
models -- such as the string case under consideration here -- when one wishes 
to avoid getting involved in the technicalities of Lagragian perturbation 
theory. In order to be able to work in terms of ordinary Eulerian
perturbations, i.e. direct comparisons between  perturbed and unperturbed
configurations at the {\it same} fixed point in the background spacetime,
one needs to use a model that spreads out beyond the single ``physical''
worldsheet in which one is ultimately interested. What is required for our
purpose is a locally extended mathematical model constructed in terms not
just of the single ``physical'' world sheet with which one is ultimately
concerned, but also of a locally surrounding spacefilling congruence of
``fictitious'' world sheets, each of which is postulated to be separately
subject to the {\it same} mathematical equations that are obeyed by the
central ``physical'' world sheet within it. 

Since it can be seen from (2.15) that the space configuration of an 
equilibrium configuration is arbitrary provided (2.19) is satisfied, it is 
easy to extend any ``physical'' equilibrium configuation of a string loop to
a neighbouring congruence of equilibrium configurations simply by extending
the field $e^\mu$ by parallel propagation orthogonal to the central
``physical' world sheet, which can be seen from (2.9) to mean that its 
covariant derivative at any point on the physical worldsheet will be given
in terms of the curvature vector of the latter by
$$\nabla_{\!\mu} e^\nu=e_\mu K^\nu\, .\eqno{(3.1)}$$
Since this construction repects the stationarity condition (2.7) the
necessary integrability conditions are satisfied as a triviality: the field
$e^\mu$ set up in this way automatically commutes with the Killing vector
field $k^\mu$ so that the two vector fields together will necessarily 
generate a well defined congruence of stationary world sheets. To complete
the construction of the ``unperturbed'' state of the fictitious string
congruence we postulate that the equilibrium conditions (2.16) to (2.19) are
satisfied on each separate world sheet.

Under such conditions all that is needed for the equilibrium state of 
the extended model to be specified completely is the choice of the running 
velocity $v$ on each separate world sheet. The simplest possibility, 
consistent with the logitudinal uniformity condition (2.18) is to take the
running velocity and, hence, the intrinsic state of each string (as specified 
by its tension $T$, for example)  to be uniform throughout. However, in 
order to retain the freedom of making alternative choices (for the purpose of 
checking consistency as a precaution against algebraic error) we shall
provisionally retain the option of letting $v$ vary from one world sheet to 
another as an arbitrary differentiable field.

Having chosen an equilibrium state in the manner just described, we now 
consider the dynamical evolution of an arbitrary small perturbation of it. 
Using the symbol $\delta$ to indicate an Eulerian differential, we can write 
the perturbations of the longitudinal dynamical equations (1.15) and (1.16)
directly as
$$\delta({\cal E}^{\rho\sigma})\nabla_{\!\rho}(\nu v_\sigma) +
{\cal E}^{\rho\sigma}\nabla_{\!\rho}\delta(\nu v_\sigma)=0
\, ,\eqno{(3.2)}$$ 
and            
$$\delta({\cal E}^{\mu\nu})\nabla_{\!\rho}(\mu u_\sigma) +
{\cal E}^{\mu\nu}\nabla_{\!\rho}\delta(\mu v_\sigma)=0
\, .\eqno{(3.3)}$$ 
After simplification by the use of the unperturbed equilibrium equations the
corresponding perturbation of the trensverse dynamical equation (1.17) takes
the form
$$\perp^{\!\mu}_{\,\rho}\big\{\delta(Uu^\nu)\nabla_{\!\nu} v^\rho -           
\delta(Tv^\nu)\nabla_{\!\nu} u^\rho+ Uu^\nu\nabla_{\!\nu}\delta(v^\rho)-
 Tv^\nu\nabla_{\!\nu}\delta(u^\rho)\, .\eqno{(3.4)}$$
Finally as the (very modest) price to be paid for the convenience of working
in terms of simple Eulerian perturbations of unrestricted fields, we must
complete the dynamic system by including the geometric surface integrability
condition (1.8), whose satisfaction in this approach is not guaranteed 
automatically as it would be in a Lagrangian treatment. Taking the direct
perturbation of the explicit version (1.18) of (1.8), and simplifying by the
use of the equilibrium equations in the same way as was done for (3.4), one 
obtains the corresponding geometric surface integrability condition for the 
perturbed field variables in the form
$$\perp^{\!\mu}_{\,\rho}\big\{\delta(u^\nu)\nabla_{\!\nu} v^\rho -           
\delta(v^\nu)\nabla_{\!\nu} u^\rho+ u^\nu\nabla_{\!\nu}\delta(v^\rho)-
 v^\nu\nabla_{\!\nu}\delta(u^\rho)\, .\eqno{(3.5)}$$

In order to proceed it is convenient to represent the longitudinal part of
the perturbation in terms of a pair of infinitesimal scalars, $\alpha$ and
$\varepsilon$ say, and to represent the transverse part of the perturbation 
in terms of a pair of vectors, $\zeta^\mu_{+}$ and $\zeta^\mu_{-}$ say,
subject to the orthogonality conditions
$$\eta^\mu_{\ \nu}\zeta^\nu_{\pm}=0\, ,\eqno{(3.6)}$$
according to the following further specifications. Subject to the
normalisation conditions (1.1) the most general perturbation of the frame
vectors $u^\mu$ and $v^\mu$ is determined by and determines the vectors 
$\zeta^\mu_{\pm}$ and the scalar $\varepsilon$ by the relations
$$\delta u^\mu=\varepsilon v^\mu + v(\zeta^\mu_{-}+\zeta^\mu_{+})\, ,
\hskip 1 cm \delta v^\mu=\varepsilon u^\mu + v(\zeta^\mu_{-}-\zeta^\mu_{+})
\, ,\eqno{(3.7)}$$
where the coefficient $v$ is the extrinsic propagation velocity as given
by (2.19) which is included in order for $\zeta^\mu_{-}$ and $\zeta^\mu_{+}$
to be interpretable as the respectively ``left'' (relatively backward)
and ``right'' (relatively forward) moving parts of the extrinsic
perturbation. Finally the other infinitesimal amplitude $\alpha$ is
defined as the logarithmic variation of the number density $\nu$ that was 
chosen in Section 1 as the basic independent variable for characterising the
intrinsic equation of state,  i.e. we set
$${\delta\nu\over\nu}=\alpha\, .\eqno{(3.8)}$$
The corresponding formula for the variation of the chemical potential or
effective mass variable $\mu$ that is required for the evaluation of (3.3) 
will then be obtainable using the equation of state as
$${\delta\mu\over\mu}=c^2 \alpha\, ,\eqno{(3.9)}$$
where $c$ is the longitudinal characteristic speed as given by the formula
(2.22). As remarked above, in the constant trace model used in most early and
many more recent discussions [22-25] of cosmic string dynamics the
longitudinal perturbation speed is that of light, i.e. $c=1$. In the more
accurate models derived by Peter[13] its value is not just less than that
of light but even less than the transverse characteristic speed, i.e. one
has $c<v<1$. On the other hand, in the low tension limit that has been 
experimentally familiar in the nonrelativistic regime since the time of
Newton and Hooke, one has $v\ll c\ll 1$.

Using condition (3.1) and the fact that the orthogonality property (3.6)
enables the contracted surface derivatives of the extrinsic perturbation
vectors to be expressed in the form
$$\tilde\nabla_{\!\mu}\zeta^\mu_\pm=-K_\mu\zeta^\mu_\pm\, ,\eqno{(3.10)}$$
we can rewrite the pair of longituninal perturbation equation (3.2) and
(3.4) in characteristic form, in terms of the quantities that have just been
introduced, as the pair of equations
$$(u^\rho\pm cv^\rho)\nabla_\rho(\varepsilon\pm c\alpha)
={(v\mp c)^2\over 2v\nu}\{(v\pm c)\zeta^\rho_{-}+(v\mp c)\zeta^\rho_{+}\}
\nabla_{\!\rho}\nu 
+2v\gamma^2(v\pm c) K_\rho\zeta^\rho_{+}\, . 
\eqno{(3.11)}$$
The pair of extrinsic equations can similarly be combined as a forward
characteristic equation,
$$  2 v\perp^{\!\sigma}_{\,\rho}(u^\nu+ v v^\nu)\nabla_{\!\nu}
\zeta^\rho_{+}=-K^\sigma\{2 v\varepsilon +(v^2+c^2)\alpha)\}
\, ,\eqno{(3.12)}$$
in which it may be remarked that $\zeta^\rho_{-}$ does not appear at all,
together with the integrability condition itself which takes the form
$$ \perp^{\!\sigma}_{\,\rho}(u^\nu- v v^\nu)\nabla_{\!\nu}
\zeta^\rho_{-}=\perp^{\!\sigma}_{\,\rho}(u^\nu+ v v^\nu)\nabla_{\!\nu}
\zeta^\rho_{+}\, .\eqno{(3.13)}$$

It can be seen that the longitudinal equations (3.11) will greatly simplify
if we now fix the choice of the unperturbed  ``fictitious'' string states of
the extended system by requiring that the intrinsic state as characterised
by $\nu$ should be uniform throughout, i.e.
$$\nabla_{\!\rho}\nu=0\, ,\eqno{(3.14)}$$
not just longitudinally along each individual world sheet as required by 
(2.18). In that case (3.11) and (3.12) will form a decoupled system of 
equations for the two longitudinal variables $\alpha$ and $\varepsilon$ and
the two independent components of the foreward moving part $\zeta^\mu_{+}$
of the extrinsic perturbation. The remaining equation (3.13) is the only
one involving the other two independent components, namely those of the
relatively backward mpving part $\zeta^\mu_{-}$ of the extrinsic 
perturbation. Application of the theorem mentionned at the end of Section 2
to the effect that the relatively backward moving extrinsic perturbation
should have zero propagation speed relative to the static Killing vector 
$k^\mu$ can be seen directly from the possibility of rewriting (3.13) in
the form
$$\perp^{\!\sigma}_{\,\rho}k^\nu\nabla_{\!\nu}\zeta^\rho_{-}=
\perp^{\!\sigma}_{\,\rho}\gamma^2(1+v^2)( k^\nu+v_{+} e^\nu)\nabla_{\!\nu}
\zeta^\rho_{+}\, ,\eqno{(3.15)}$$
where $v_{+}$ is the foreward extrinsic propagation speed given by (2.21).
The decoupled system for the other four independent components is given by
the analogous version of the characteristic evolution equation (3.12) for
the foreward moving part of the extrinsic perturbation, i.e.
$$2\gamma v (1+v^2)\perp^{\!\sigma}_{\,\rho}(k^\nu+v_{+} e^\nu)
\nabla_{\!\nu}\zeta^\rho_{+}= -K^\sigma\{2 v\varepsilon + (v^2+c^2)
\alpha\}\, ,\eqno{(3.16)}$$
together with the longitudinal equations which are expressible as
$$ (1+v^2)(k^\nu+v_{+}e^\nu)\nabla_{\!\nu}\varepsilon
+\{ v(1+c^2)k^\nu+(v^2+c^2) e^\nu\}\nabla_{\!\nu}\alpha
=4\gamma v^2 K_\rho\zeta^\rho_{+}\, ,\eqno{(3.17)}$$
and 
$$ k^\nu\nabla_{\!\nu}\varepsilon-\gamma^2\{v(1-c^2)k^\nu+(v^2-c^2)
e^\nu\}\nabla_{\!\nu}\alpha=0\, .\eqno{(3.18)}$$

\bigskip\noindent
{\bf 4. Global solution of the perturbation equations 
in the circular case.}
\medskip

In order to obtain a system whose global solution is easily calculable,
we now restrict ourselves to the case in which the unperturbed equilibrium 
state is a circular ring configuration with fixed radius, $r$ say, which
means that the tangential derivative of the curvature vector $K^\mu$
will be given by
$$\eta^\nu_{\ \rho}\nabla_{\!\nu} K^\mu=-K_\nu K^\nu e_\rho e^\mu
\, ,\eqno{(4.1)}$$
where its magnitude has the fixed value given by
$$K^\mu K_\mu=1/r^2\, .\eqno{(4.2)}$$

For such a configuration the perturbation may be conveniently analysed in
terms of the natural time and angle coordinates which are specified (modulo 
a choice of origin) on the equilibrium world sheet by
$$k^\nu\nabla_{\!\nu} t=1\, ,\ \ \ k^\nu\nabla_{\!\nu} \theta=0\, ,\ \ \
e^\nu\nabla_{\!\nu} t =0\, ,\ \ \ e^\nu\nabla_{\!\nu} \theta=1/r
\, .\eqno{(4.3)}$$
We can consider a quite general perturbation as a discrete sum of Fourier
components with respect to the periodic coordinate $\theta$. For the scalar
longitudinal perturbation variables $\alpha$ and $\varepsilon$ the Fourier
components are definable for each (by convention positive) integer 
$n=0, 1, 2,$ simply as the real parts of corresponding complex variables
having the form
$$ \alpha=\check\alpha\, {\rm e}^{-i n\theta}\, ,\hskip 1 cm
 \varepsilon=\check\varepsilon\, {\rm  e}^{-i n\theta}\, ,\eqno{(4.4)}$$
where $\check\alpha$ and $\check\varepsilon$ satisfy the rotational 
invariance condition
$$ e^\nu\nabla_{\!\nu}\check\alpha=0\, ,\hskip 1 cm
 e^\nu\nabla_{\!\nu}\check\varepsilon=0\, ,\eqno{(4.5)}$$
so that they are functions only of $t$ but not of $\theta$.
The corresponding Fourier components of the extrinsic displacement vectors
$\zeta_\pm^\mu$ will be similarly expressible in the form
$$\zeta_\pm^\mu=\check\zeta_\pm^\mu\,{\rm  e}^{-i n\theta}\, ,\eqno{(4.6)}$$
where the complex amplitude vectors $\check\zeta_\pm^\mu $ are rotationally 
invariant in the appropriate sense, which is not that of covariant 
differentiation but that of Lie differentiation with respect to the 
relevant angle Killing vector, $re^\mu$. This Lie type constancy condition
is just the natural generalisation of the condition (4.1), satisfied by the 
curvature vector $K^\mu$ itself, being expressible in terms of 
straightforeward covariant differentiation as
$$\eta^\nu_{\ \rho}\nabla_{\!\nu}\,\check\zeta_\pm^{\,\mu}=-K_\nu\,
\check\zeta_\pm^{\,\nu} e_\rho e^\mu\, .\eqno{(4.7)}$$

Condition (2.3) ensures that with respect to the time Killing vector 
$k^\mu$ there will be no distinction between covariant and Lie 
differentiation which we shall denote simply by a dot. Thus for any Fourier
mode of the kind just described, it can be seen that the derivations in the
perturbation equations obtained in the previous section can be evaluated using the simple scalar expressions,
$$ k^\nu\nabla_{\!\nu}\,\alpha=\dot\alpha\, ,\ \ \
 k^\nu\nabla_{\!\nu}\,\varepsilon=\dot\varepsilon\, ,\ \ \
e^\nu\nabla_{\!\nu}\,\alpha=-i{n\over r}\alpha\, ,\ \ \
e^\nu\nabla_{\!\nu}\,\varepsilon=-i{n\over r}\varepsilon\, ,\eqno{(4.8)}$$
together with the corresponding vectorial expressions,
$$ k^\nu\nabla_{\!\nu}\zeta_\pm^\mu=\dot\zeta_\pm^\mu\, ,\hskip 1 cm
\perp^{\!\mu}_{\,\rho}e^\nu\nabla_{\!\nu}\,\zeta_\pm^\rho=-i{n\over r}
\zeta_\pm^\mu\, .\eqno{(4.9)}$$

Substitution of these relations enables us to reduce Eq. (3.15) for the
relatively backward moving part $\zeta_{-}^\mu$ of the extrinsic 
perturbation to the ordinary differential relation
$$ (1-v^2)\dot\zeta_{-}^\mu=(1+v^2)\zeta_{+}^\mu-{2inv\over r}\zeta_{+}^\mu
\, . \eqno{(4.10)}$$

The decoupled system (3.16) - (3.18) for the remaining variables $\alpha$,
$\varepsilon$, and the two independent components of $\zeta_{+}^\mu$ is
analogously reducible to the ordinary differential system,
$$ 2\gamma v\Big( (1+v^2)\dot\zeta_{+}^\mu+ {2inv\over r}\zeta_{+}^\mu
\Big)=-K^\mu\big(2v\varepsilon+(v^2+c^2)\alpha\big)\, ,\eqno{(4.11)}$$
$$(1+v^2)\dot\varepsilon-{2inv\over r}\varepsilon+v(1+c^2)\dot\alpha
-(v^2+c^2){in\over r}\alpha=2\beta\, ,\eqno{(4.12)}$$
$$(1-v^2)\dot\varepsilon-v(1-c^2)\dot\alpha+(v^2-c^2){in\over r}\alpha
=O\, ,\eqno{(4.13)}$$
using the abbreviation
$$\beta=2\gamma v^2 K_\nu\,\zeta_{+}^\nu\, .\eqno{(4.14)}$$

The generic solutionof the foregoing evolution equations will have the form 
of a normal mode as given in terms of some constant real or complex
$\omega$ by
$$\check\alpha=\langle\alpha\rangle\,{\rm e}^{i\omega t}\, ,\ \ \ \
\check\varepsilon=\langle\varepsilon\rangle\,{\rm e}^{i\omega t}\, ,\ \ \ \
\check\zeta_\pm^\mu=\langle\zeta_\pm^\mu\rangle\,{\rm e}^{i\omega t}
\, ,\eqno{(4.15)}$$
where the (complex) bracket labelled quantities are all {\it constant} --
not just with respect to the angle but also with respect to time, i.e.
$$ \langle\dot\alpha\rangle=0\, , \hskip 1 cm
 \langle\dot\alpha\rangle=0\, , \hskip 1 cm
\langle\dot\zeta_\pm^\mu\rangle=0\, .\eqno{(4.16)}$$

It can be seen from (4.10) that, as was to be expected from the general 
equilibrium theory described in section 2, for all values of $n$ there are 
effectively {\it static} (zero eigenfrequency) perturbation modes for
arbitrary time independent complex amplitudes $\check\zeta_{-}^\mu$ of
the relatively backward moving extrinsic displacement vector with
vanishing values of the relatively foreward part and of the longitudinal
perturbation amplitudes, i.e. modes characterised by
$$\omega=0\, ,\ \ \ \zeta_{-}^\mu= \langle\zeta_{-}^\mu\rangle\, 
{\rm e}^{-in\theta}\,\ \ \ \zeta_{+}^\mu=0\, \ \ \ \alpha=\varepsilon=0\, .
\eqno{(4.17)}$$

It can also be seen that the azimuthal component of $\zeta_{+}^\mu$ as
projected orthogonally to the curvature (i.e. parallel to the axis of 
symmetry of the ring) will {\it decouple} completely from the rest of the
system, which shows that for each value of $n$ there will be a
corresponding relatively foreward propagating extrinsic mode with 
displacement parallel to the axis of symmetry that takes the normal mode
form (4.15) with
$$\omega =n\Omega_{+}\, ,\ \ \ \zeta_{-}^\mu=0\, ,\ \ \ \beta=0\, \ \ \
\alpha=\varepsilon=0\, ,\eqno{(4.18)}$$
using a notation scheme in which $\Omega$ and $\Omega_{+}$ denote the
angular velocities corresponding to the unperturbed ring rotation velocity
$v$ as given by (2.19) and the relatively foreward rotating extrinsic 
velocity $v_{+}$ as given by (2.21), i.e.
$$\Omega={v\over r}\, ,\hskip 1 cm\Omega_{+}={v_{+}\over r}
\, ,\eqno{4.19)}$$
so that (4.18) can be interpreted as meaning that relatively foreward moving
azimuthal modes consisting purely of extrinsic displacements parallel to the
axis of symmetry will propagate with the same (characteristic) velocity
$v_{+}$ independently of their wavelength.

As the only nontrivial (wavelength dependent) part of the proble we are thus
left with the subsystem of longitudinal and purely radial perturbations as
specified in terms of just three perturbation variables  which can be listed
as the scalar $\beta$ defined to be proportional to the radial component of 
the perturbation vector $\zeta^\mu_{+}$ by the relation (4.14), together 
with the longitudinal perturbation amplitudes $\varepsilon$ and $\alpha$.
The system governing the normal modes
$$\varepsilon=\langle\varepsilon\rangle\,{\rm e}^{i(\omega t-n\theta)}\, , 
\hskip 1 cm \alpha=\langle\alpha\rangle\,{\rm e}^{i(\omega t-n\theta)}\, ,
\hskip 1 cm \beta=\langle\beta\rangle\,{\rm e}^{i(\omega t-n\theta)}\, ,
\ \eqno{(4.20)}$$
of these three remaining variables is expressible in the matrix form
$$ \left(\matrix{2 & v^2+c^2  & (1\!+\!v^2)\sigma\!-\!2n\cr
(1\!+\!v^2)\sigma\!-\!2n & v^2(c^2\!+\!1)\sigma\!-\!(v^2\!+\!c^2)n & 2
\cr (1\!-\!v^2)\sigma &  v^2(c^2\!-\!1)\sigma\!+\!(v^2\!-\!c^2)n & 0}
\right) \left(\matrix{v\varepsilon\cr \alpha \cr i\beta r}\right)
= \left(\matrix{0\cr 0\cr 0}\right) \eqno{(4.21)}$$
in which the perturbation frequency variable $\omega$ has been replaced by
the corresponding dimensionless ratio
$$\sigma=\omega/\Omega\, .$$
The system (4.21) evidently provides an eigenvalue equation for the frequency
variable $\sigma$, which must be such that the corresponding determinant 
vanishes, i.e.
$$ \matrix{v^2(1\!+\!v^2)(1\!-\!c^2 v^2)\sigma^3 + 2 v^2\big(c^2\!-\!v^2\!-\!
2(1\!-\!c^2 v^2)\big) n\sigma^2\hskip 3 cm \cr
+\!\big(4v^2(1\!-\!c^2)(n^2\!-\!1)\!-\!(1\!+\!v^2)(c^2\!-\!v^2)(n^2\!+\!1)
\big)\sigma 
+2(c^2\!-\!v^2)(n^2\!-\!1)n=0\, .} \eqno{(4.23)}$$

The condition for strict dynamic stability of the mode specified by a given
value of the integer $n$ is that the corresponding eigenvalue equation
(4.23) should have three real roots, distinct from each other -- and from 
the zero eigenvalue $\omega=0$ corresponding to the solution (4.17) -- in 
which case there will exist corresponding real values for the triplet of
amplitudes $\langle\varepsilon\rangle$, $\langle\alpha\rangle$, and
$i\langle\beta\rangle$. The presence of the factor $i$ in the last of
these is interpretable as meaning that there is a phase lag whereby the
maximum variation of the intrinsic variables (such as the tension $T$)
as determined by $\alpha$, and of the longitudinal velocity variation, as
determined by $\varepsilon$, will always coincide with the phase at which
the radial displacement as determined by $\beta$ is zero. The existence
of complex roots will evidently imply strict dynamical instability, i.e.
a mode that grows exponentially, while the special case of zero or
mutually coincident eigenvalue solutions of (4.23) corresponds to the
existence of a mode that is only marginally stable, not strictly stable, 
so that the normal mode form (4.20) may not be sufficient to provide the
complete solution of its time evolutionas in the generic case. It is evident 
that (4.23) has a zero eigensolution only for the trivial case $n=0$ of a
nearby concentric circular equilibrium state and for the almost equally 
trivial case $n=1$, for which there are marginally stable state that grow 
not exponentially but {\it linearly} in time, and that are interpretable
merely as equilibrium states with respect to nearby, slightly Lorentz 
boosted frames.

\bigskip\noindent
{\bf 5. Demonstrations of stability in particular cases.}
\medskip

It is easy to check that, subject to the standard conditions [8, 11] for
microscopic stability and causality, i.e.
$$0< v^2\leq 1\, ,\hskip 1 cm 0< c^2\leq 1\, ,\eqno{(5.1)}$$
the conditions for dynamical stability will always be satisfied for the
lowest modes that are characterised by $n=0$ and $n=1$.

In the case $n=0$, corresponding to modes of pure conformal expansion or
contraction in which the ring conserves its circular shape, the odd powers 
of $\sigma$ in (4.23) drop out, so that the equation is easily solved to
give either the trivial possibility $\sigma=0$ -- corresponding to a 
perturbation in which the original ring solution is merely replaced by 
another neighbouring ring state -- or else the fundamental oscilation 
frequency which can easily be seen to be given by
$$\sigma^2={c^2-3c^2 v^2+3 v^2-v^4\over v^2(1+v^2)(1-c^2 v^2)}\, .
\eqno{5.2}$$
In the first general study of circular equilibrium states of the kind we
are considering it was argued that stability with respect to axisymmetry
preserving perturbatins could be derived from the condition that the 
relevant total mass energy ${\cal M}$ should be a minimum. That requirement
can be seen to be precisely equivalent to the reality requirement that must
be satisfied by the dimensionless frequency $\sigma$ since the formula that
was obtained [8, 9] for the second derivative of ${\cal M}$ can be seen to 
be related to the fundamental angular frequency $\omega=\Omega\sigma$
obtained from (5.2) by the very simple relation
$$\omega^2={1\over{\cal M}}{{\rm d}^2{\cal M}\over{\rm d} r^2}
\, .\eqno{(5.3)}$$
It was remarked in the original discussion that the right hand side of
(5.3) would be positive, ensuring stability of the fundamental mode, in all
the most obvious applications. A closer scrutiny shows that this is actually
an understatement: the right hand side of (5.2) will in fact be positive 
in all cases without exception, provided the inequalities (5.1) are not
violated, as can be seen by regrouping the terms in the form
$$\sigma^2={2v^2(1-c^2)+(c^2+v^2)(1-v^2)\over v^2(1+v^2)(1-c^2v^2)}
\, .\eqno{(5.4)}$$

When we go on to consider the next level of excitation as characterised
by $n=1$ we find that since the inhomogenious term in (4.23) still drops
out we are again left with a trivial {\it zero} eigenfrequency (which 
corresponds to the possibility of a simple lateral displacement and whose 
degeneracy with (4.17) signals the existence of the marginally stable
linearly growing mode corresponding to the further trivial possibility of
a simple Lorentz boost), together with the pair of non-trivial oscillation
frequencies that are obtained as solutions of the quadratic equation
$$v^2(1\!+\!v^2)(1\!-\!c^2v^2)\sigma^2 + 2 v^2\big(c^2\!-\!v^2\!
-\!2(1\!-\!c^2v^2)\big)\sigma-2(c^2\!-\!v^2)(1\!+\!v^2)=0\, .\eqno{(5.5)}$$
By casting the explicit solution in the form
$$\sigma={v\big(v^2\!-\!c^2\!+\!2(1\!-c^2v^2)\big)
\pm\sqrt{c^2(1\!-\!v^4)\!+\!v^2(1\!-\!c^2v^2)}
\sqrt{(1\!-\!v^4)+(1\!-\!c^2v^2)}
\over v(1+v^2)(1-c^2v^2)}\eqno{(5.6)}$$
it can be seen that, as in the previous case, the roots will always be real 
so long as the inequalities (5.1) are satisfied, since all the bracketted 
combinations within the square roots are manifestly positive.

For the purpose of investigating higher modes it is useful to use a rescaled
variable
$$x={1+v^2\over 2n}\sigma\, ,\eqno{(5.7)}$$
and to work in terms of a dimensionless ``distension'' parameter,
$$\Delta={c^2-v^2\over v_{+}^2(1-c^2v^2)}\, ,\eqno{(5.8)}$$
which can be seen to play a critical discriminating role. In terms of this
parameter $\Delta$ and of the prograde rotation  speed $v_{+}$ as given by 
(2.21) instead of the original parameters $c$ and $v$, the eigenvalue 
equation (4.23) can be rewritten in the convenient form
$$\Big(x-1-{1\over n}\Big)\Big(x-1+{1\over n}\Big)=\Delta(1-v_{+}^{\, 2})
\Big(x-1+{1\over n^2}\Big){x\over x+\Delta}\, ,\eqno{(5.9)}$$
whose solution can be interpreted as the geometric problem of finding the
intersections of the parameter independent parabola determined by the left 
hand side with the hyperbola determined by the righthand side. Noting that 
the vertical asymptote to the hyperbola occurs where $x=-\Delta$, it can 
easily be seen that there will be a real negative root (corresponding to a 
retrograde mode) if and only if the distension parameter $\Delta$ is 
positive, and that if this condition is satisfied there will also be two 
real positive roots (corresponding to prograde modes). The condition of
positivity of $\Delta$ is evidently equivalent to the condition that the
dimensionless ratio
$$\chi={c\over v}\eqno{(5.10)}$$
of longitudinal to extrinsic characteristic speed should be greater
that unity,
$$\chi >1\, .\eqno{(5.11)}$$

This sufficient condition for stability can obviously be extended to cover 
the case for which the discriminant vanishes, i.e. for which
$$ \chi=1\, .\eqno{5.12)}$$
This special condition automatically holds in the special case of the
integrable ``constant product'' model with equation of state $UT=m^4$ for
some fixed mass parameter $m$ -- which arises naturally from Kaluza Klein 
theory [8] and is also relevant as a macroscopic (effectively renormalised)
 ``warm'' string approximation allowing [20, 21] for a thermal or other 
distribution of microscopic wiggles on an underlying Goto - Nambu model -- 
and which is characterised by the condition that the two kinds of 
characteristic speed remain exactly equal as the tension varies. Since this
property allows the equations of motion to be easily integrated explicitly
in a flat background [20] it could have been predicted in advance that the
present perturbation problem should also be easily soluble in this particular 
case. Such an expectation is indeed justified, since when the condition 
(5.12) holds the constant term (4.23) evidently drops out so that again, as
for the other cases $n=0$ and $n=1$ one obtains a zero eigenfrequency
$\omega=0$ while the other two eigenfrequencies are got by solving a 
quadratic which leads to a result that is expressible very simply as
$$\omega=(n\pm 1)\Omega_{+}\, ,\eqno{(5.13)}$$
where $\Omega_{+}$ is the angular velocity of relatively foreward 
characteristic motion as given by (4.19). The manifest reality of this result
confirms that for this particular model no instability occurs for any mode.

Knowing that the conditions (5.1) ensure stability in the large $n$ limit 
[11] and having found no possibility of instability in the lowest modes,
$n=0$ and $n=1$, nor for any value of $n$ when $\chi\geq1$, one might be
tempted to conjecture that all modes will necessarily be stable. To show that,
despite the existence of all these positive results, unstable counterexamples
{\it can} occur for models having equilibrium states with smaller values of 
the longitudinal to extrinsic propagation velocity,$\chi<1$, it is
sufficient to consider the Newtonian limit
$$v^2\ll 1\, ,\hskip 1 cm c^2\ll 1\, .\eqno{(5.14)}$$
for which the general eigenvalue equation (4.23) reduces to a form in which
the ratio $\chi$ is the only parameter involved appart from the ratio $n$,
the result that is obtained being expressible as
$$\sigma^3-4n\sigma^2+\Big(4(n^2\!-\!1)-(n^2\!+\!1)(\chi^2-1)\big)\sigma
+2n(n^2\!-\!1)(\chi^2-1)=0\, .\eqno{(5.15)}$$ 
The results obtained above to the effect that all the roots are real for
$n=0$ and $n=1$ will still of course apply to this limit in particular, but 
as soon as we go on to consider the next level of excitation as given by
$n=2$ we find that dynamic instability will indeed occur for any state that
is ``too taut'' in the sense of having too small a value of $\chi$. For the
equation (5.15) the stipulation that the roots be real is equivalent to the
requirement that the ratio $\chi$ be greater than a small but stricty
positive critical limit value $\chi_c$ say, i.e. the condition for
stability will be satisfied if and only if
$$\chi\geq \chi_c\, ,\eqno{(5.16)}$$
where the critical value is about $1/2$, being given approximately by
$$\chi_c\simeq 0.47\, .\eqno{(5.17)}$$
Since it is only displacements in the plane of the undisturbed
configuration that are involved, this limit should be easy to test
experimentally by spinning up an ordinary elastic string loop such as a
thin elastic band on a smooth plane surface. In the Hookean low tension
limit, $\chi$ will be very large so the danger of instability will not
arise, but as the angular velocity and therefore also the tension of the
loop is increased one would expect $\chi$ to decrease monotonically and
ultimately to fall below the instability threshold (5.17).

In the relativistic regime relevant to the superconducting string loop 
problem that originally motivated this work the situation is, of course,
more complex, depending not just on the ratio $\chi$ but on the absolute
values of $v$ and $c$. A more thorough analytic and numerical exploration 
of the zones of stability and instability in the general case, and of their
relationship to the predictions of particular superconducting cosmic string 
models, will be postponed for a subsequent article [26].

\vfill\eject

\noindent
{\bf References.}
\medskip
\parindent=0 cm

[1] R.L. Davis, E.P.S. Shellard, `The physics of vortex 
superconductivity. 2', {\it Phys. Lett.} {\bf B209} (1988) 485.
\smallskip

[2] R.L. Davis, `Semitopological solitons',
{\it Phys. Rev.} {\bf D38} (1988) 3722.
\smallskip

[3] R.L. Davis, E.P.S. Shellard, `Cosmic vortons',
{\it Nucl. Phys.} {\bf B323} (1989) 209.
\smallskip

[4] J. Ostriker, C. Thompson, E. Witten, `Cosmological effects of
 superconducting strings', {\it Phys. Lett.} {\bf 180} (1986) 231.
\smallskip

[5] E. Copeland, M. Hindmarsh, N. Turok, `Cosmic springs',
{\it Phys. Rev. Lett.} {\bf 56} (1987) 910.
\smallskip

[6] D. Haws, M. Hindmarsh, N. Turok, `Superconducting strings or 
springs?' {\it Phys. Lett.} {\bf B209} (1988) 255.
\smallskip

[7] E. Copeland, D. Haws, M. Hindmarsh, N. Turok, `Dynamics of and 
radiation from superconducting strings and springs',
{\it Nucl. Phys.} {\bf B306} (1988) 908.
\smallskip

[8] B. Carter,
`Covariant mechanics of simple and conducting strings and membranes',
in {\it The Formation and Evolution of Cosmic Strings}, ed.G. Gibbons,
S. Hawking. T. Vachaspati (Cambridge. U. P., 1990) 143-178.
\smallskip

[9] B. Carter,  `Mechanics of Cosmic Rings',
{\it Phys. Lett.} {\bf B238} (1990) 166.
\smallskip

[10] B. Carter, `Duality Relation between Charged Elastic Strings and 
Superconducting Cosmic Strings', {\it Phys. Lett.} {\bf B224} (1989) 61.
\smallskip

[11] B. Carter, `Stability and Characteristic Propagation  Speeds in 
Superconducting Cosmic and other String Models',
{\it Phys. Lett.} {\bf B229} (1989) 446.
\smallskip

[12] B. Carter, `Basic Brane Theory',
{\it Class. Quantum Grav.} {\bf 9} (1992) 19.
\smallskip

[13] P. Peter, `Superconducting cosmic string: equation of state for 
spacelike and timelike current in the neutral limit',
{\it Phys. Rev.} {\bf D45} (1992) 1091.
\smallskip

[14] P. Peter, `Influence of the electric coupling strenght in current 
carrying cosmic strings',
{\it Phys. Rev.} {\bf D46} (1992) 3335.
\smallskip

[15] P. Peter, `Electromagnetically supported cosmic string loops',
{\it Phys. Lett.} {\bf 298} (1993) 60.
\smallskip

[16] B. Carter, `Cosmic Rings as a Chump Dark Matter Candidate?',
in {\it Particle astrophysics: Early Universe and Cosmic Structures}, 
ed. J-M. Alimi, A. Blanchard, A. Bouquet, F. Martin de Volnay, 
J. Tran Thanh Van, (Editions Fronti\`eres, Gif-sur-Yvette, 1990) 213-221.
\smallskip

[17] B. Carter, `Cosmological Relic Distribution of Conducting String 
Loops', {\it Ann. N.Y. Acad. Sci.} {\bf 647} (1991) 758.
\smallskip

[18] B. Carter, V.P. Frolov, O Heinrich, `Mechanics of Stationary 
Strings: Separability for non-dispersive models in black hole 
background', {\it Class. Quantum Grav.} {\bf 8} (1991) 135.
\smallskip

[19] B. Carter `Basic Brane Mechanics',  in {\it Relativistic 
Astrophysics and Gravitation} ed. S. Gottloeber, J.P. Muecket, 
V. Mueller,  (World Scientific, Singapore, 1992) 300-319.
\smallskip

[20] B. Carter, `Integrable Equation of State for Noisy Cosmic String',
{\it Phys. Rev.} {\bf D41} (1990) 3869.
\smallskip

[21] A. Vilenkin, `Effect of small scale structure on the dynamics of
cosmic strings', {\it Phys. Rev.} {\bf D41} (1990) 3038.
\smallskip

[22] D.N. Spergel, T. Piran, J. Goodman, `Dynamics of superconducting 
strings', {\it Nucl. Phys.} {\bf B291} (1987) 847.
\smallskip

[23] A. Vilenkin, T. Vachaspati, `Electromagnetic radiation from 
superconducting strings', {\it Phys. Rev.} {\bf D39} (1989) 379
\smallskip

[24] D.N. Spergel, W.H. Press, R.J. Scherrer, `Electromagnetic 
selfinteraction of superconducting cosmic strings',
{\it Phys. Rev.} {\bf D39} (1989) 379.
\smallskip

[25] P. Amsterdamski, `Evolution of superconducting cosmic loops',
{\it Phys. Rev.} {\bf D39} (1989) 1534.
\smallskip

[26] X. Martin, `Zones of dynamical instability for rotating string
loops', {\it Phys. Rev.} {\bf D50} (1994) 682.

\end